# A POSSIBLE MECHANISM OF THE KIRKWOOD GAP FORMATIONS AT THE VERY BEGINNING

Kazantsev A. M.
Astronomical Observatory of Taras Shevchenko National University of Kyiv
ankaz@knu.ua

**ABSTRACT**

The orbits of asteroids from the MPC catalogue of May 31, 2020 with absolute magnitudes $H < 16^m$, in the 3:1, 5:2 and 2:1 mean motion resonances (MMRs) with Jupiter were selected. The number of the orbits in the 2:1 MMR is dozens times greater than in the other two resonances.

There are fragments of parent bodies of neighbour asteroid families, in particular the Themis family, among bodies in the 2:1 MMR.

Numerical calculations were performed to evaluate the evolution of the selected orbits over hundreds of thousand years. Perturbations from all eight major planets and the relativistic effects of orbital perihelion displacement were taken into account. For all orbits in the 3:1 and 5:2 MMRs an increase in the orbit eccentricities, which are sufficient for the bodies to approach Mars, was obtained. In the 2:1 MMR, a sufficient increase in the orbit eccentricities was not detected.

An increase in orbit eccentricities in this resonance can occur due to the action of non-gravitational effects (NGEs). The action of the Yarkovsky effect can explain the exit of an asteroid with a size of 5 km from the 2:1 MMR over a period about 1 billion years or more.

More than 2 billion years ago, there were dozens of bodies over 50 km in size in the 2:1 gap. To form the gap in the 2:1 resonance at the very beginning, the physical conditions in the asteroid belt had to be significantly different from the today ones. In particular, the intensity of the solar radiation in the early Solar system could be much higher as compared to the today one.

## 1. INTRODUCTION

The existence of gaps in the main belt of asteroids (MBA) was discovered by Kirkwood more than 150 years ago. These gaps exist in the vicinity of those values of the semimajor axes of the orbits $a_c$, where the mean motion of the asteroid $n_a$ and the mean motion of Jupiter $n_J$ correlate as integers

$$n_a/n_J = p/q \qquad (1)$$

This ratio is called mean motion resonance. Semimajor axes of small body orbits periodically oscillate in regard to the $a_c$ value, in the proximity of the commensurability. Therefore for all orbits in commensurability, equation (1) holds only approximately.

Over 150 years, a large quantity of studies devoted for asteroid orbits in different resonances with Jupiter were published. A number of features of the evolution of such orbits have been identified. In particular, it was found that the MMRs are sources of near-Earth asteroids (Bottke et al. 2002).

However, to date no general mechanism of the origin of the gaps has been developed. In some papers (Jefferys 1967) the mutual collisions of asteroids in resonances was proposed as such a mechanism. Other authors believed that asteroids simply did not form in the resonances by sticking together of planetosimals, in contradistinction to zones outside the resonances (Heppenheimer 1980). Such mechanisms have not found a support from other scientists.

The most common point of view of the origin of the gaps is that asteroids go out of a resonance after approaching the major planets, in particular Mars. For example, Wisdom (1982) indicates on the possibility of increase in the orbit eccentricities in the 3:1 MMR ($a_c$ = 2,500 AU) from $e$ = 0.05 to $e$ > 0.3. This $e$ values are sufficient for the bodies to approach Mars. This conclusion was confirmed in Moons (1996). Besides a possibility of a noticeable increase in the eccentricities of the orbits in the 5:2 MMR ($a_c$ = 2,828 AU) were showed (Yoshikawa 1991; Moons 1996). However, in the 2:1 MMR ($a_c$ = 3,278 AU) a significant increase in the orbits eccentricities was not detected in these papers and in Lemaitre & Henrard (1990) as well.

Such a possibility was shown in Chrenko et al. (2015). In this paper, in addition to the gravitational influence of the planets, the non-gravitational Yarkovsky effect was taken into account. It was found in the paper that the exit time from the gap of bodies up to 5 km in size should be of from several hundreds of million years to 1 billion years. The exit from the gap of large bodies was not considered.

Thus, the mechanism of origin of the gaps in the asteroid belt remains an incompletely solved problem to date. In this study it is made an attempt to propose such a mechanism.

## 2. SELECTION OF THE RESONANT ORBITS

The main results in the study have been obtained on the basis of numerical integration of the orbit evolutions of the Solar system bodies. A detailed description of the integration method is found in Kazantsev (2002). Perturbations from all eight major planets (Mercury – Neptune) were taken into account in the present calculations. The relativistic effects of orbital perihelion displacement were taken into account as well. All asteroids were treated as massless particles.

The calculations error for the semimajor axes of the asteroid orbits is $1\times10^{-7}$ AU, whereas for the eccentricities it constitutes $1\times10^{-7}$ at intervals up to 50,000 yr. This accuracy is maintained for different MBAs in the absence of approaches to the planets.

To more fully description the problem of the gaps it is desirable to consider the distributions of orbits in vicinity of the 3:1, 5:2 and 2:1 MMRs. The most noticeable gaps in the MBA are located there. There is a resonant zone for each commensurability, i.e. a region in the $a$ - $e$ coordinates, in which orbits can still be in the resonance. Usually the resonant zone has a shape of an inverted trapezium.

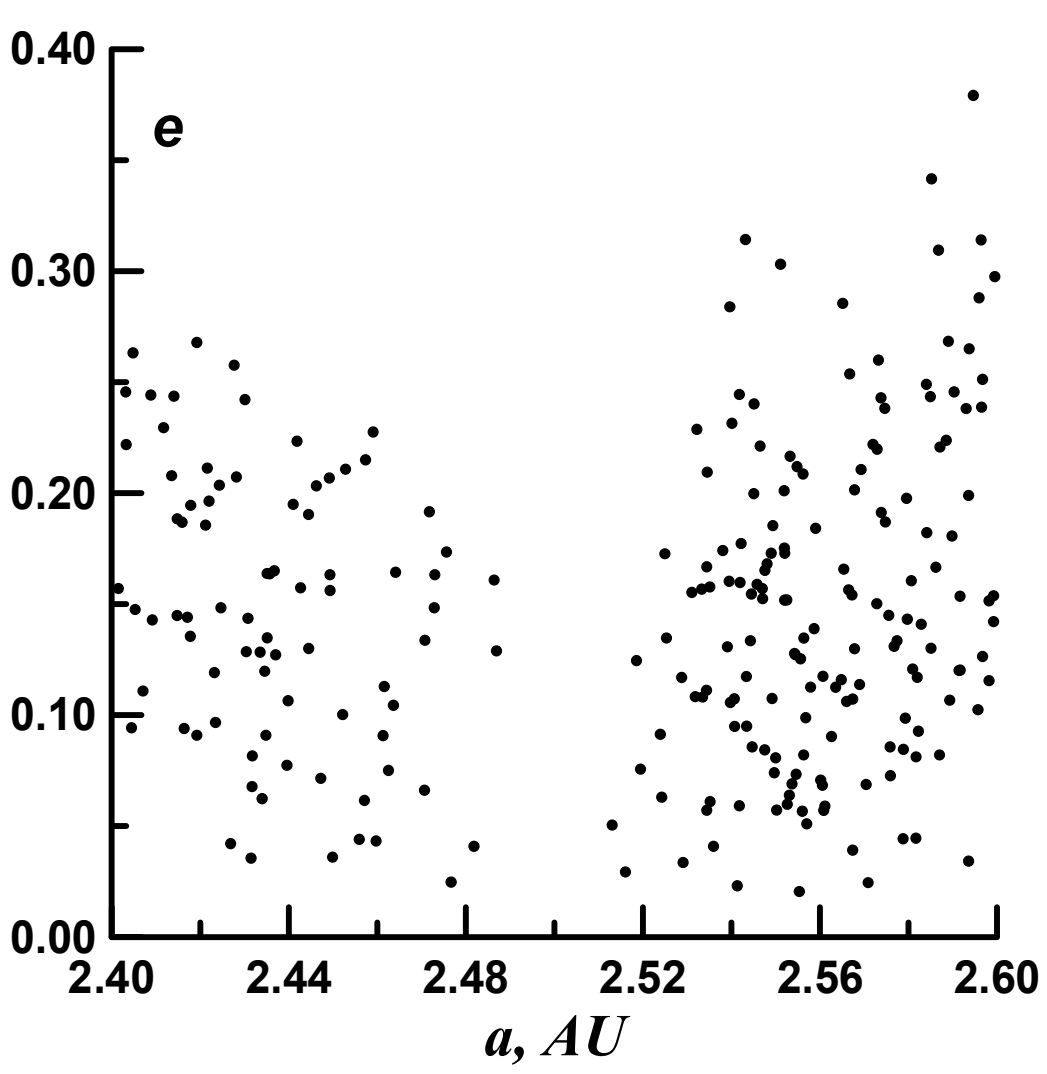

***Figure 1.*** The $a$ – $e$ distribution of orbits of asteroids with $H < 12^m$ in the vicinity of the 3:1 resonance

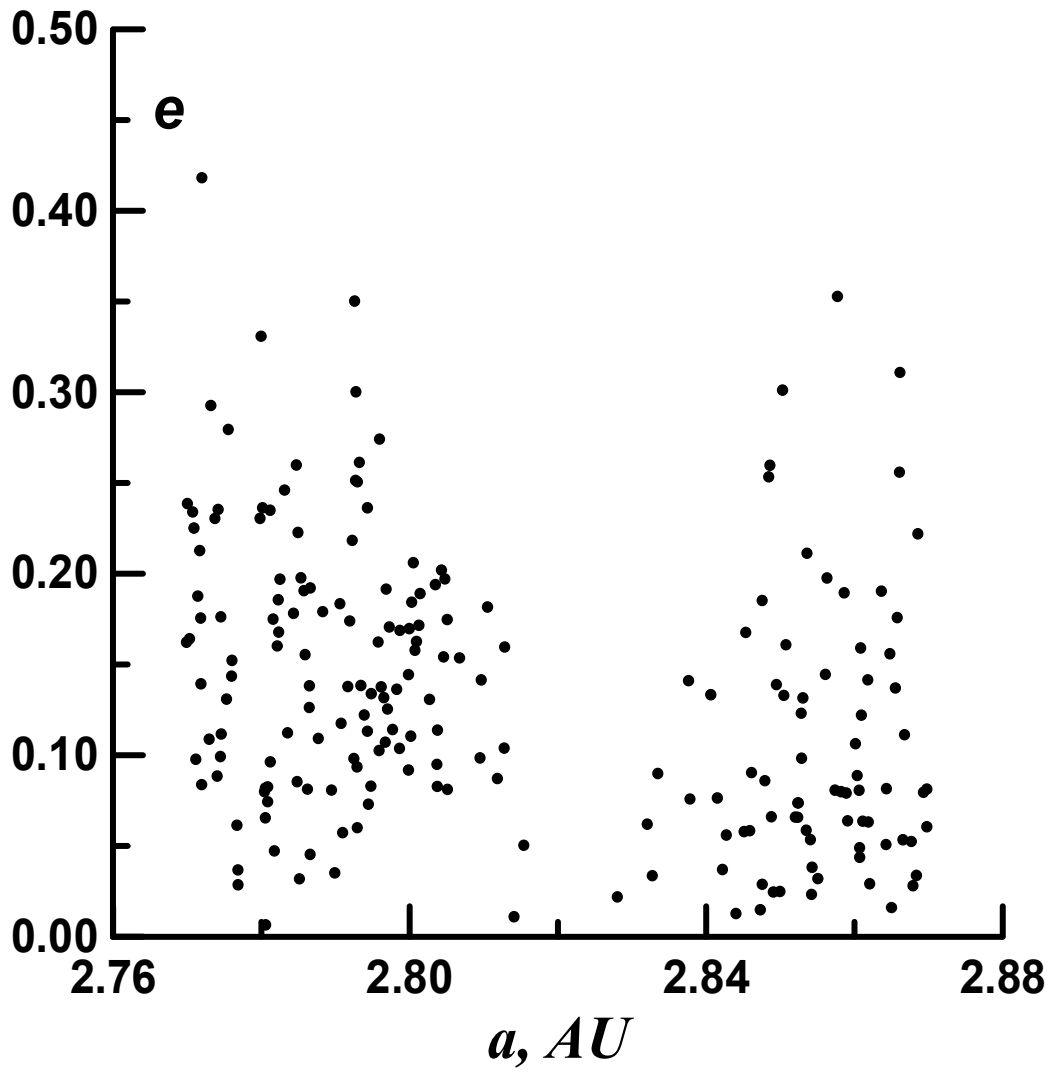

***Figure 2.*** The $a$ – $e$ distribution of orbits of asteroids with $H < 12^m$ in the vicinity of the 5:2 resonance

Figs 1-3 show the distributions of the asteroid orbits in the $a$ – $e$ coordinates in the vicinity of resonances 3: 1, 5: 2 and 2: 1. The orbits of bodies with absolute magnitudes $H < 12^m$ are pointed here. The orbits from the MPC catalogue of January, 21, 2022 were selected.

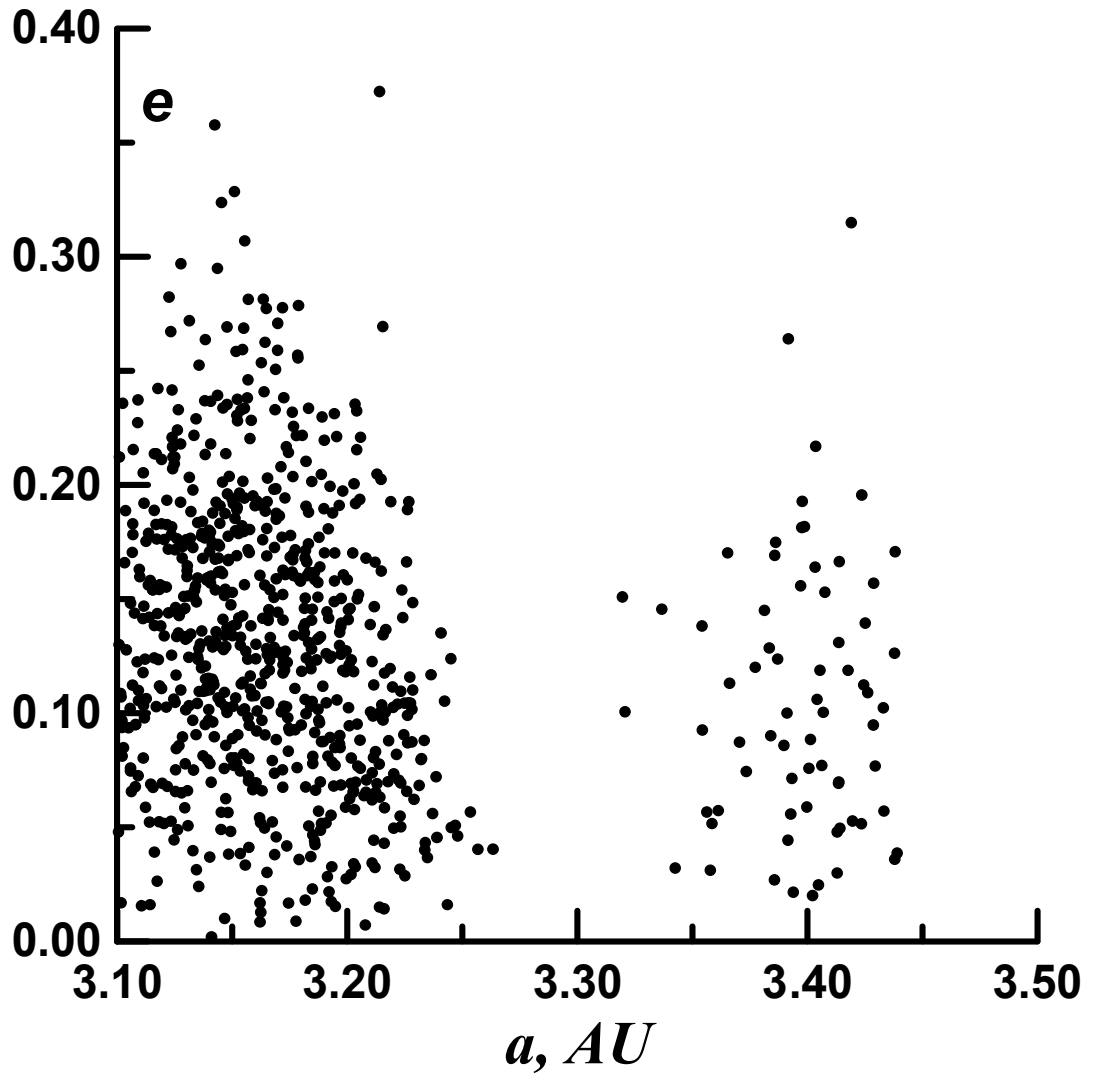

***Figure 3.*** The $a$ – $e$ distribution of orbits of asteroids with $H < 12^m$ in the vicinity of the 2:1 resonance

The figures clearly show the fact of existence of the gaps in the vicinity of the resonances. However, these distributions do not indicate the complete absence of resonant asteroids in these gaps. Such asteroids exist, but they have usually smaller sizes (greater the $H$ values).

All resonant asteroids with $H < 16^m$ were selected in these three commensurabilities. Initially the boundaries of the resonant zones in the $a – e$ coordinates were determined. For this purpose, numerical calculations of the evolution of model resonant orbits with three initial values of eccentricities (0.05, 0.15, and 0.30) were performed over relatively short intervals (up to 2000 yr). During this time, the semimajor axes of the orbits already perform several oscillations in regard to the $a_c$ value. This provides a possibility to determine the ranges of change of the $a$ values at different $e$. The amplitudes of oscillations of the $a$ values depend on the initial values of the mean anomaly $M$. The boundaries of the resonant zones correspond to the maximum amplitudes of oscillations of the $a$ values.

Fig. 4 shows the course of the evolution of orbits in the $a – e$ coordinates in the 5:2 MMR for different initial values of $e$ and $M$ (solid lines). The initial value of the inclination $i$ was $5^o$ for all orbits. At larger inclinations, the amplitude of oscillations of the $a$ value decreases. Therefore, the widest resonant zones were determined according to the calculations (dotted lines in Fig. 4). The boundaries of the resonance zones in the 3:1 and 2:1 MMRs were determined in the same way.

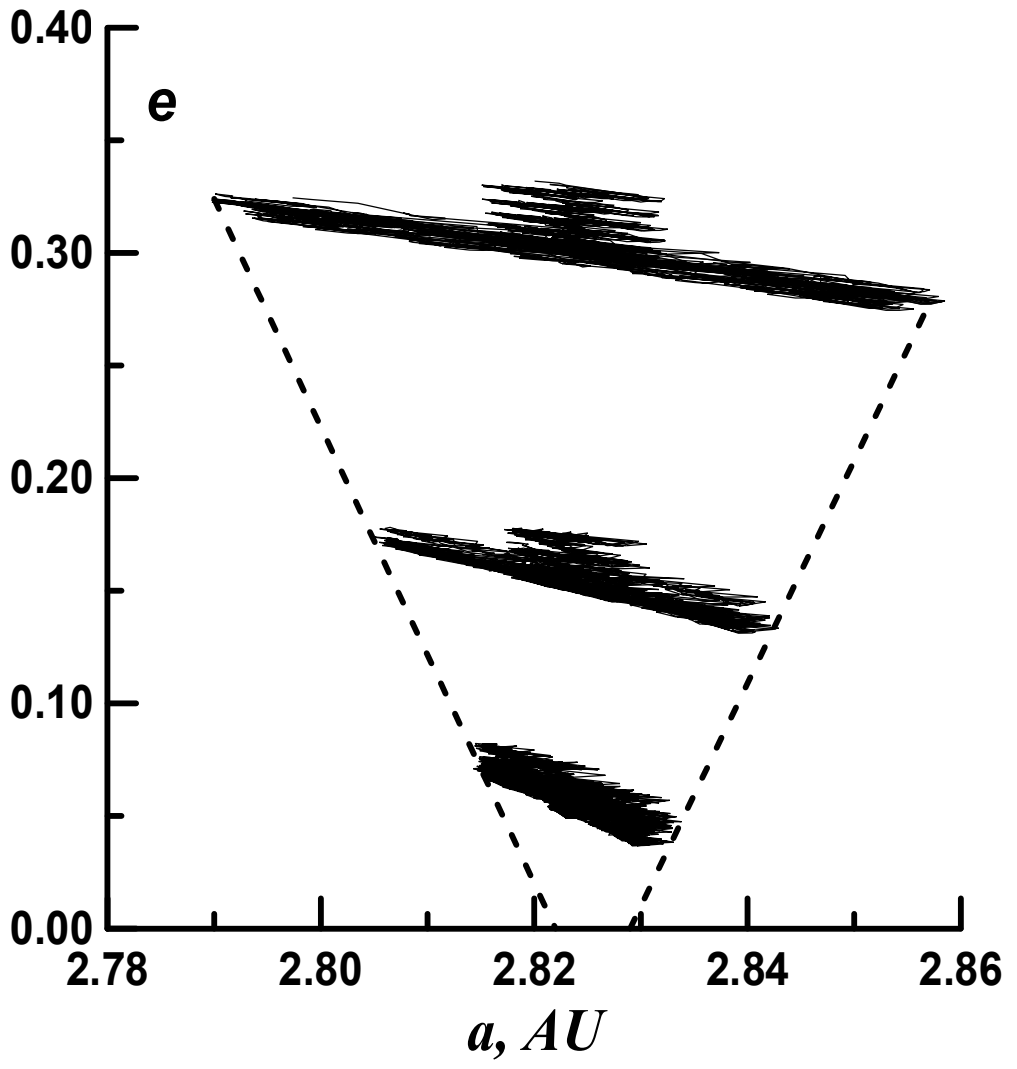

***Figure 4.*** The evolution of orbits in the $a$ – $e$ coordinates in the 5:2 resonance

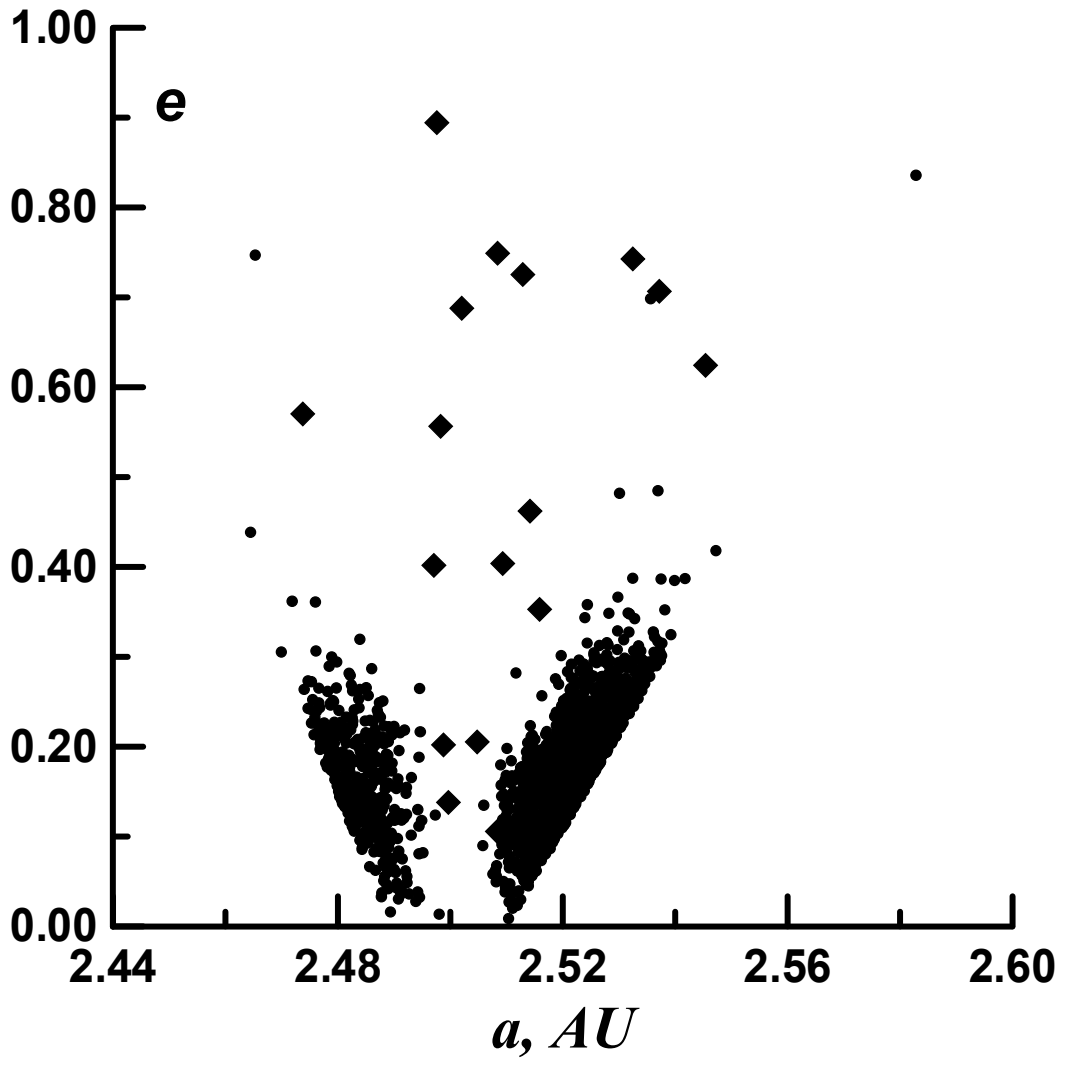

***Figure 5.*** The resonant zone with the resonant orbits (squares) in the 3:1 commensurability

Then all orbits with the $a$ and $e$ values within the resonant zones and with $H < 16^m$ from the MPC catalogue of January 21, 2022 were selected. In the 3:1 commensurability the orbit number was 1262, in the 5:2 commensurability – 1309, in the 2:1 commensurability – 3307. The detection of resonant orbits in each of the arrays was based on numerical calculations of evolution over short time intervals (up to 2,000 yr). At such intervals, it is already obvious whether such an orbit is resonant or not. The semimajor axes of the resonant orbits periodically intersect the values of $a_c$ in the course of the evolution, in contrast to the non-resonant orbits.

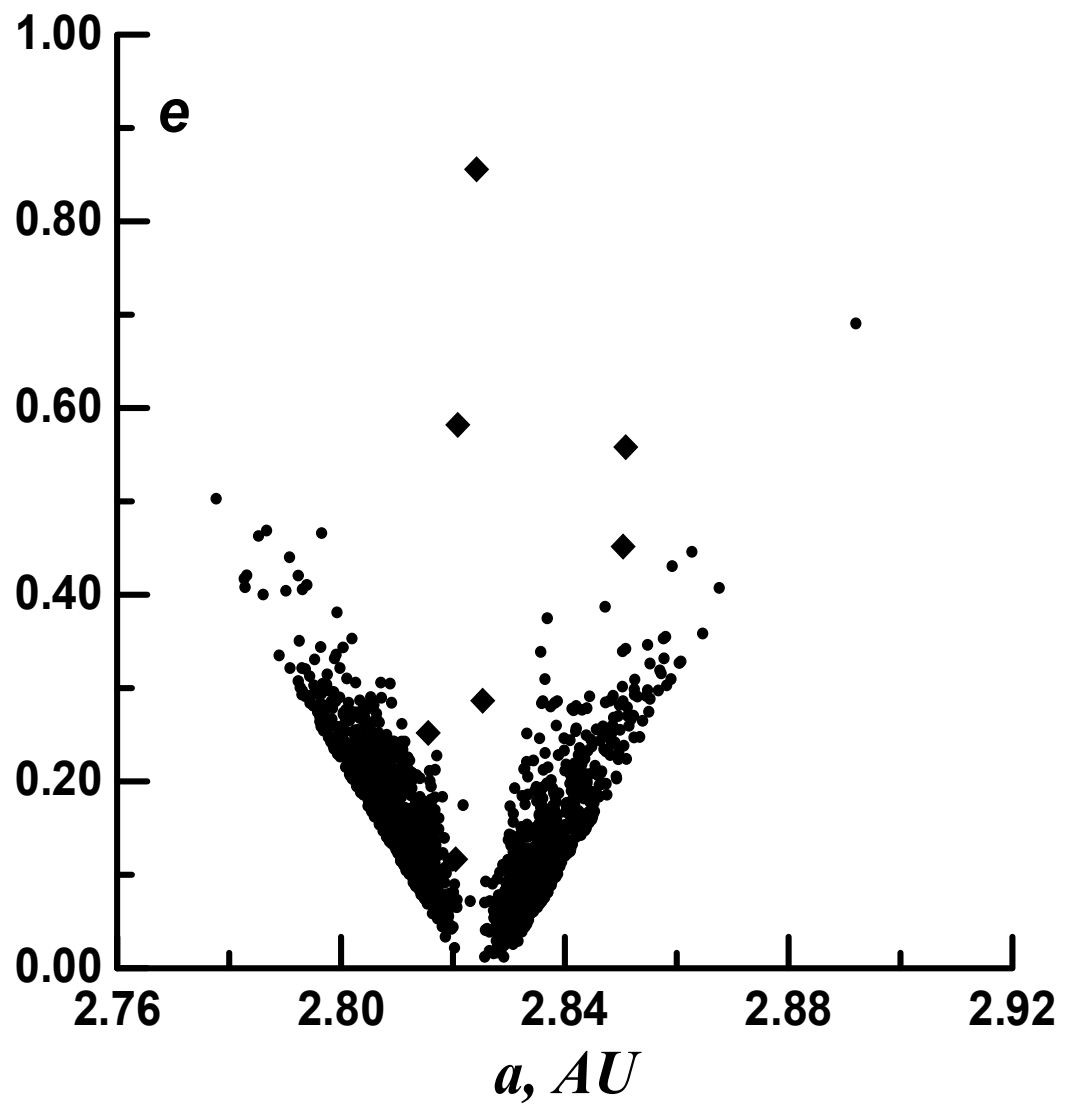

***Figure 6.*** The resonant zone with the resonant orbits (squares) in the 5:2 commensurability

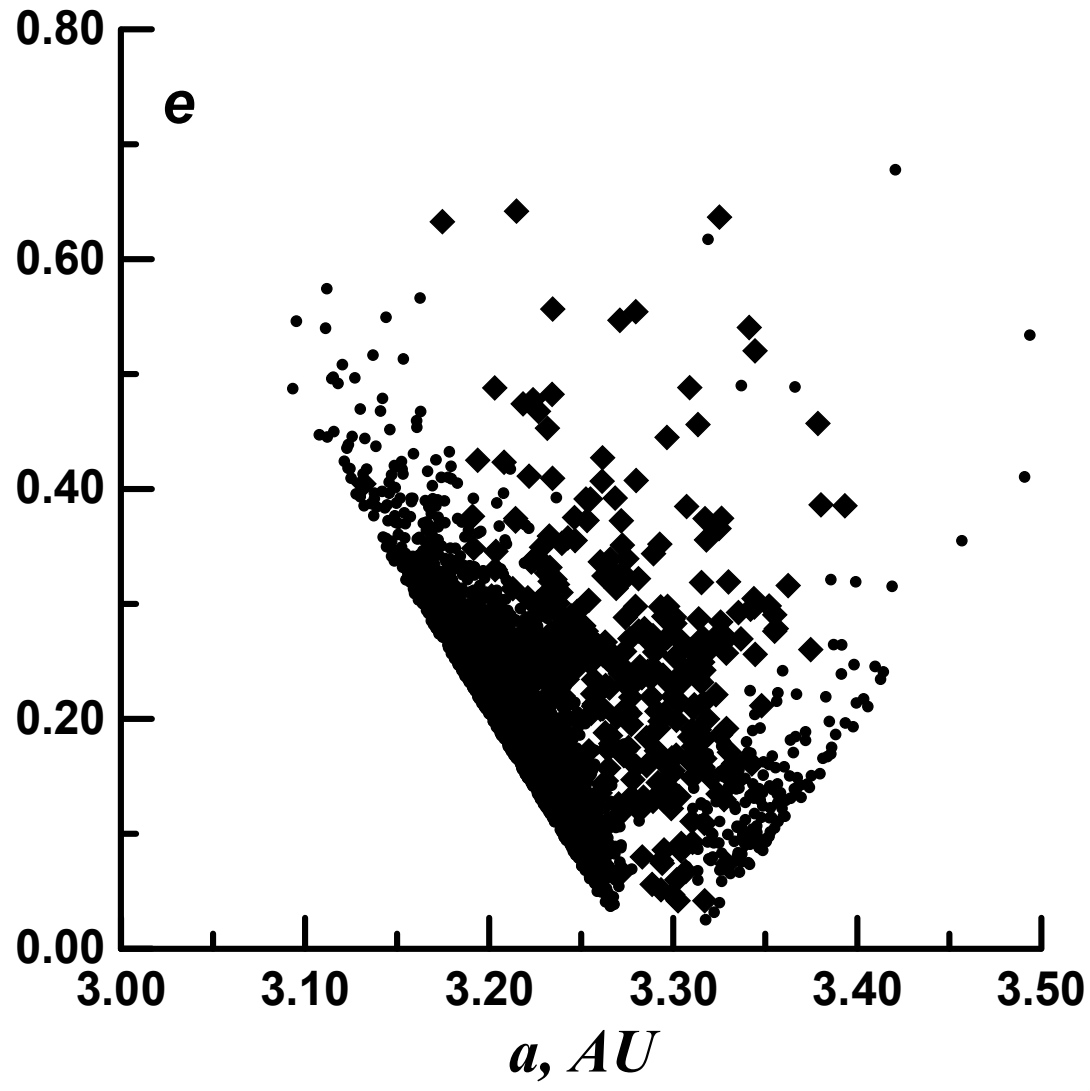

***Figure 7.*** The resonant zone with the resonant orbits (squares) in the 2:1 commensurability

In the 3:1 commensurability 17 resonant orbits were found, in the 5:2 commensurability – 7, in the 2:1 commensurability – 350. Figs 5-7 show the distributions of all selected orbits in the $a - e$ coordinates within each of those resonant zones. The resonant orbits are marked with squares.

## 3. THE MECHANISMS OF REPLENISHMENT OF THE GAPS

The above data show that the number of the resonant orbits in the 2:1 MMR is much more, than in two other ones. Moreover, the observational selection effect can only reduce the relative number of orbits in the 2:1 commensurability. It is the first peculiarity.

One can see from Figs 1-3 that in all three gaps the large bodies are completely or almost completely absent. Only in the 2:1 resonance there is one body with $H < 12^m$. It is the asteroid 1362 Griqua with $H = 11.2^m$. The minimum $H$ value of bodies in the detected resonant orbits in the 3:1 MMR is $13.8^m$, and in the 5:2 MMR – $15.3^m$. It is the second peculiarity.

A correct explanation of these two features will help to find the answer to the origin of the gaps in general.

Small bodies can fall in resonances after approaches or collisions with other asteroids, or after disintegration a large body. It is clear that this is possible for bodies in orbits which are located very close to the resonance zones. After such events, a body can almost immediately go into the gap. But such events are very rare.

Besides, small bodies can go to resonances from more remote areas, but for a much longer time. This is possible due to the Yarkovsky effect. Just the Yarkovsky effect is widely considered as the main mechanism of replenishment of the gaps. Although the Yarkovsky effect changes the orbit of a body very slowly, this mechanism acts on large quantity of bodies at once, in contrast to collisions or approaches.

As can be seen from Fig. 3, the concentration of bodies on the left of the 2:1 resonance is much greater than the concentration of bodies on the right. Therefore, one can assume that the vast majority of bodies in the resonance have went there from the left side ($a < a_c$).

There are several large asteroid families in this area, including the Themis and the Hygiea families. In particular, large asteroid families near resonances may be significant sources of replenishment of the gaps. The distributions of members of families in the coordinates of proper semimajor axis – size ($a'$ – $D$) can be evidence of this. The data can be found in Masiero et al. (2013). The paper lists 76 families with a total number of more than 38,000 bodies. In addition to the proper orbit elements of the asteroids, their WISE-based dimensions are also given. The name of each family in this paper matches with the number of the largest asteroid in the family. For example, the 008 family is known as the Flora family.

If the family members are selected correctly, the $a'$ – $D$ distribution should have a central maximum with the wings in both sides, for example, the 163 family (Fig. 8). If the orbit of the parent body of the family is located near the resonance, the $a'$ – $D$ distribution may have a truncated wing, for example, the Themis family (Fig. 9). Here the edge of the right wing of the family is truncated by the 2:1 resonance. Since the orbit elements in resonance undergo significant changes, such bodies are not considered as the family members. Another family with a truncated wing by the 2:1 resonance is the Hygiea family (Fig. 10). To more detailed picture in the $a'$ – $D$ distribution, the bodies greater than 100 km in sizes are not pointed in the figure.

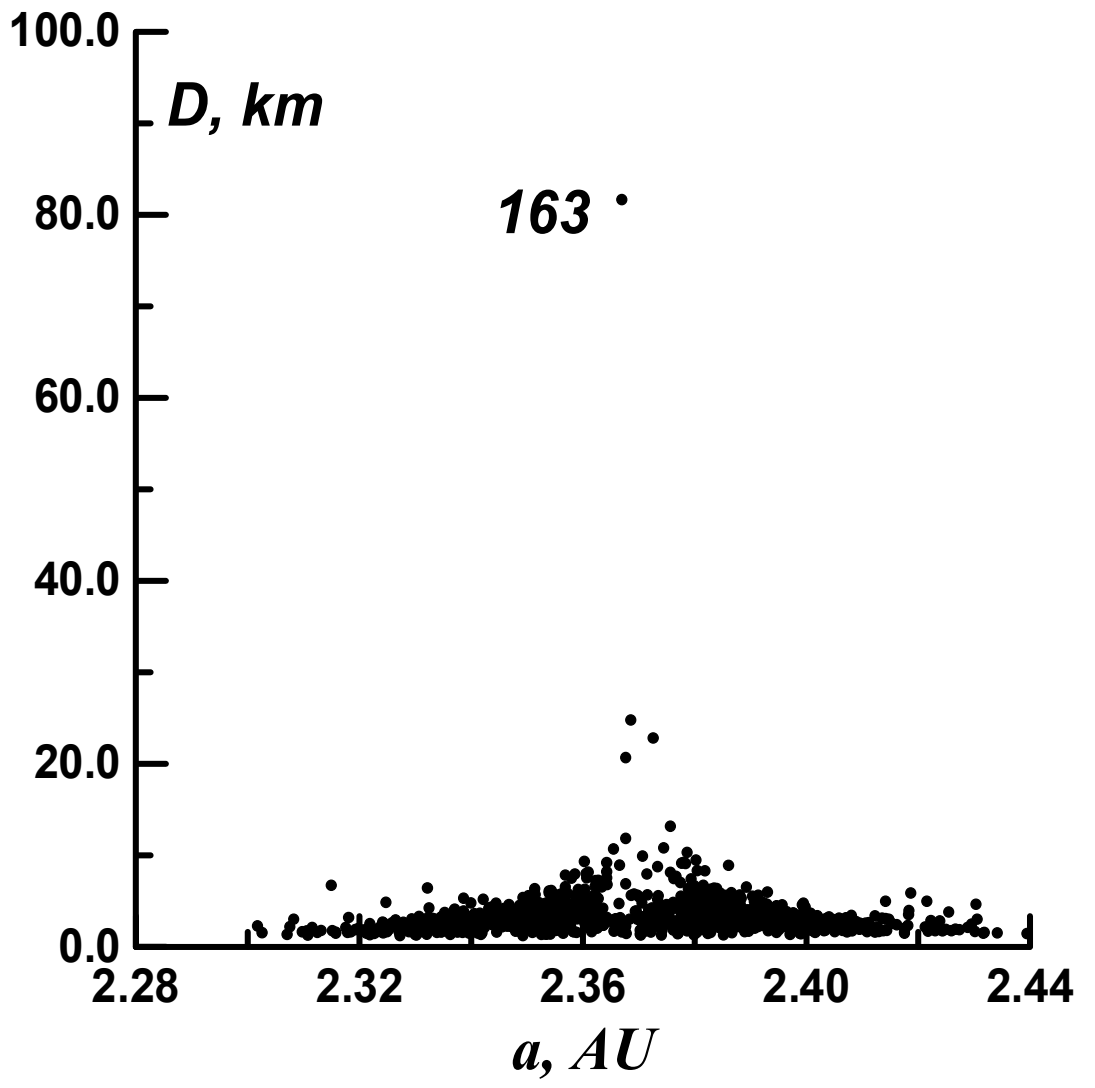

***Figure 8.*** The $a'$ – $D$ distribution of the 163 family

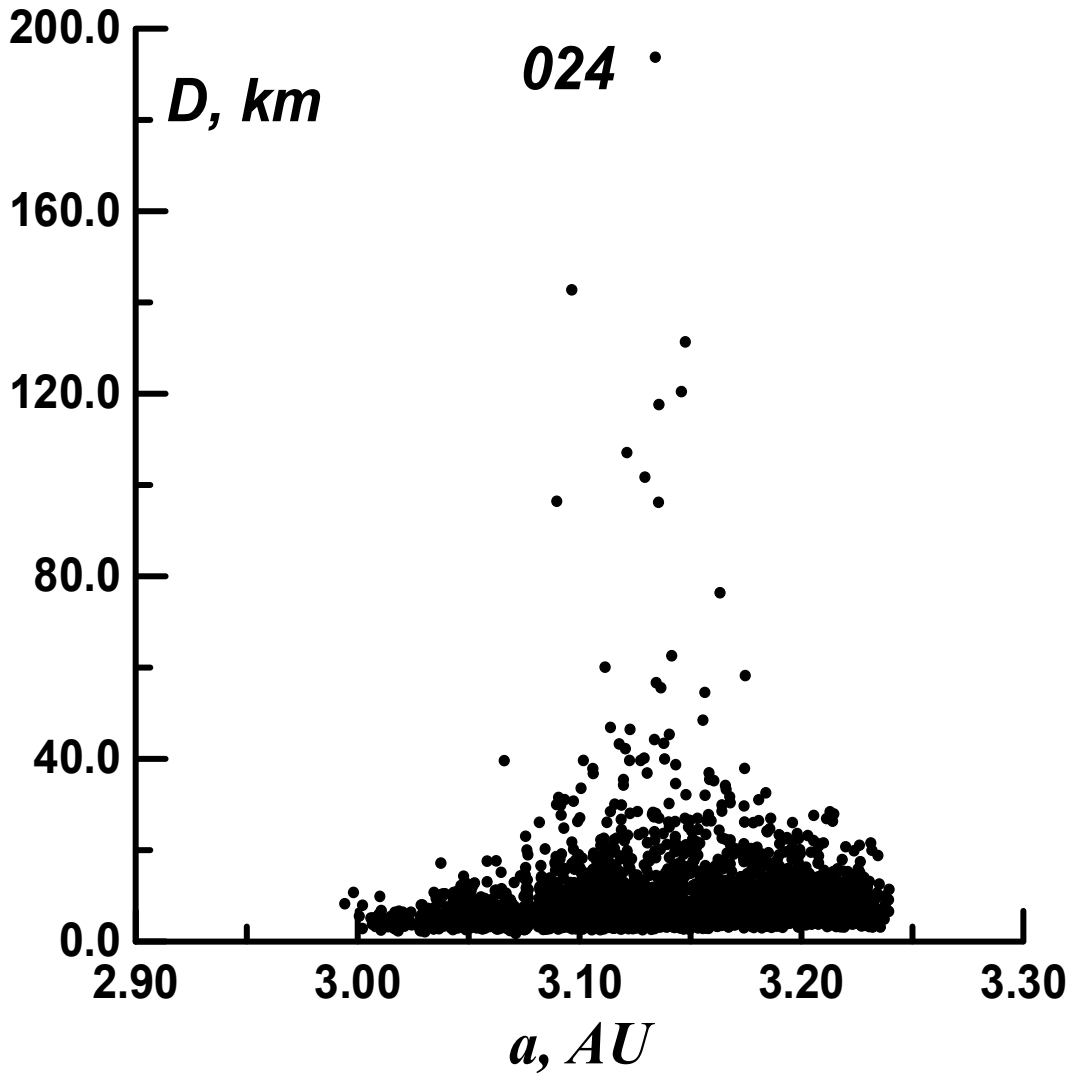

***Figure 9.*** The $a'$ – $D$ distribution of the Themis family

A specific $a'$ – $D$ distribution of a family (central maximum with wings in both sides) can be of both primary and evolutionary origin. When the parent body is destroyed, the fragments (family members) fly in different directions at certain velocities. On average, the smaller a fragment size, the greater its velocity, i.e, the greater difference between the semimajor axes of the orbits of the parent body and the fragment. However, the $D(a')$ dependences of the families is considered by scientists as the result of the Yarkovsky effect's action (Bottke et al. 2001). The action of this effect forms qualitatively the same $a'$ – $D$ distribution as the primary explosion. It is believed that the primary velocities of the fragments are insufficient to form long wings of families (Bottke et al. 2001).

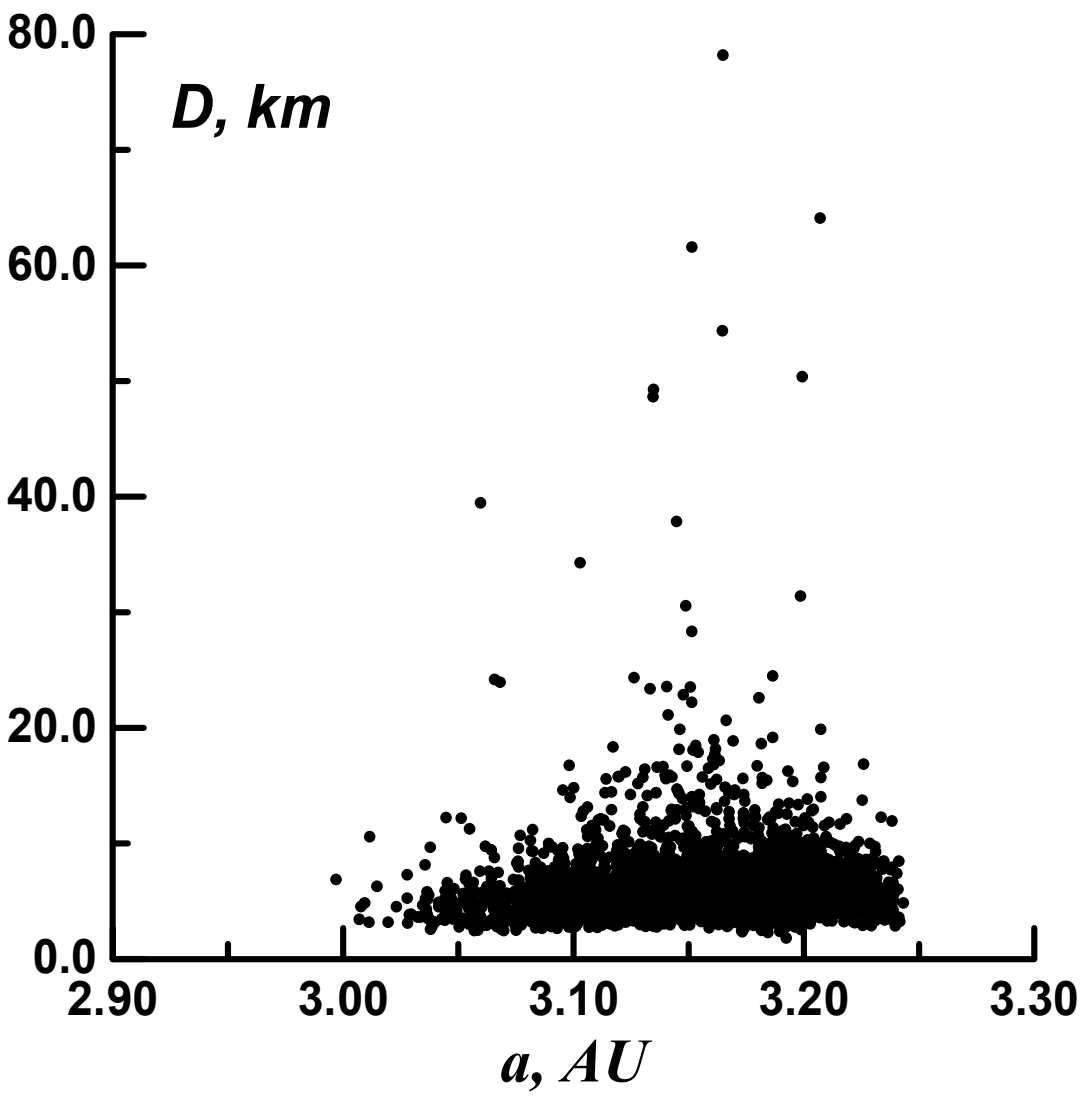

***Figure 10.*** The $a'$ – $D$ distribution of the Hygiea family

There are other arguments for the dominant influence of the Yarkovsky effect on the formation of the $D(a')$ dependences of families. In particular, in Hanuš et al. (2013), an anisotropic distribution of spin vectors of asteroids in families was revealed. Bodies with prograde rotation (β > 0°) are usually located in the right wing and bodies with retrograde rotation (β < 0°) – in the left. Just such spin vectors orientation should be in case of the family wing formations due to the Yarkovsky effect.

A reduction in the mean albedo with increasing semimajor axis for almost all correctly identified families listed in Masiero et al. (2013), is found in Kazantsev & Kazantseva (2014). It is clear that such a dependence can not be formed during the destruction of the parent body.

The $a' - D$ distributions with truncated right wings for the families Themis (Fig. 9) and Hygiea (Fig. 10) indicate that some of the fragments of the parent bodies of these families fell in the gap 2:1. It is clear the asteroids in orbits on the left of the gap, which do not belong to families (background asteroids), can also be in the gap. Therefore, among the selected 350 bodies in resonant orbits may be both asteroids from the families and the background asteroids. It is desirable to separate some asteroids from others. To do this, the distribution of the number of asteroids by inclination should be considered. The inclinations of resonant orbits in the course of the evolution vary within relatively small limits compared to other orbit elements.

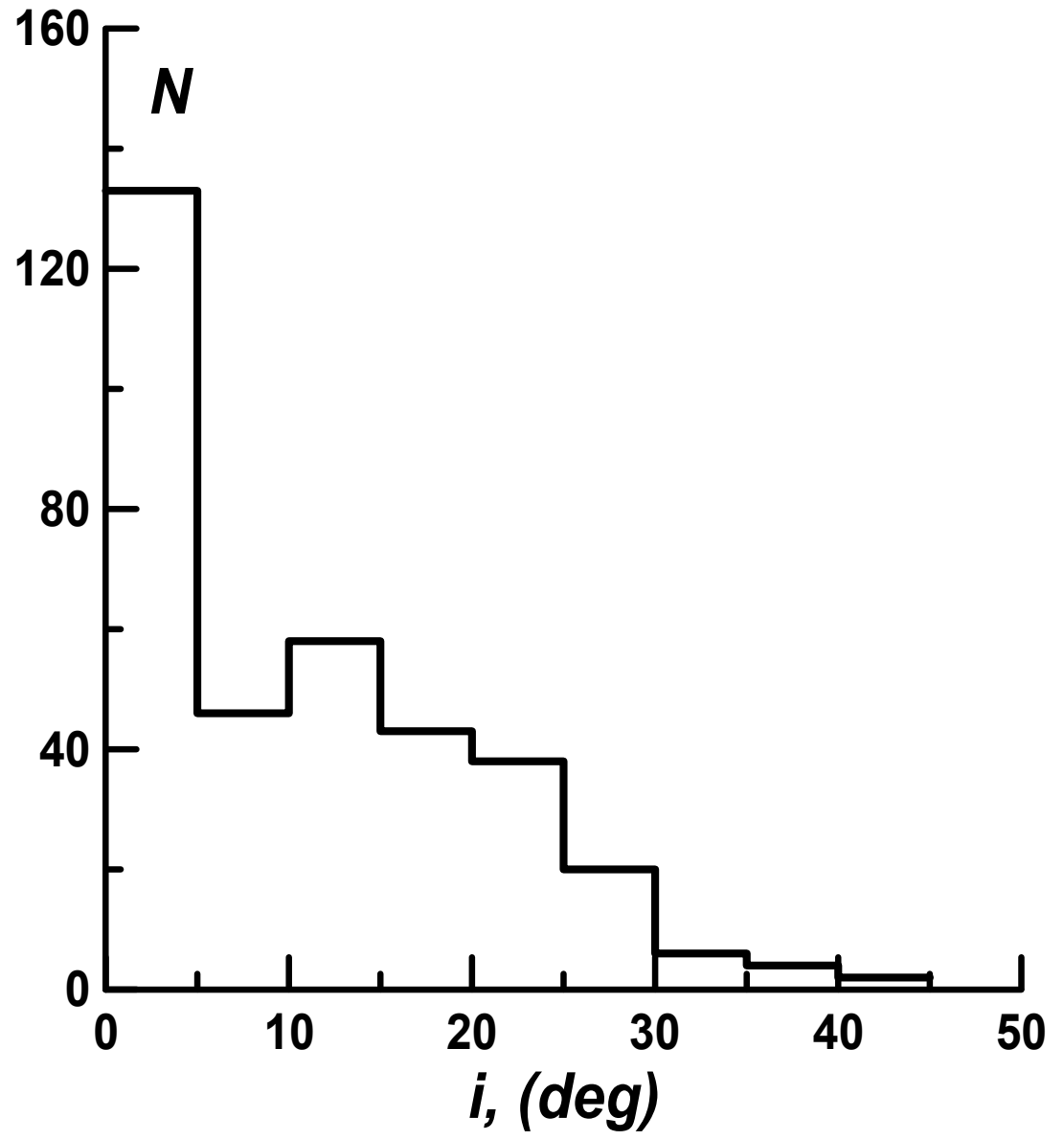

***Figure 11.*** The $N(i)$ distribution of the resonant orbits in the 2:1 gap

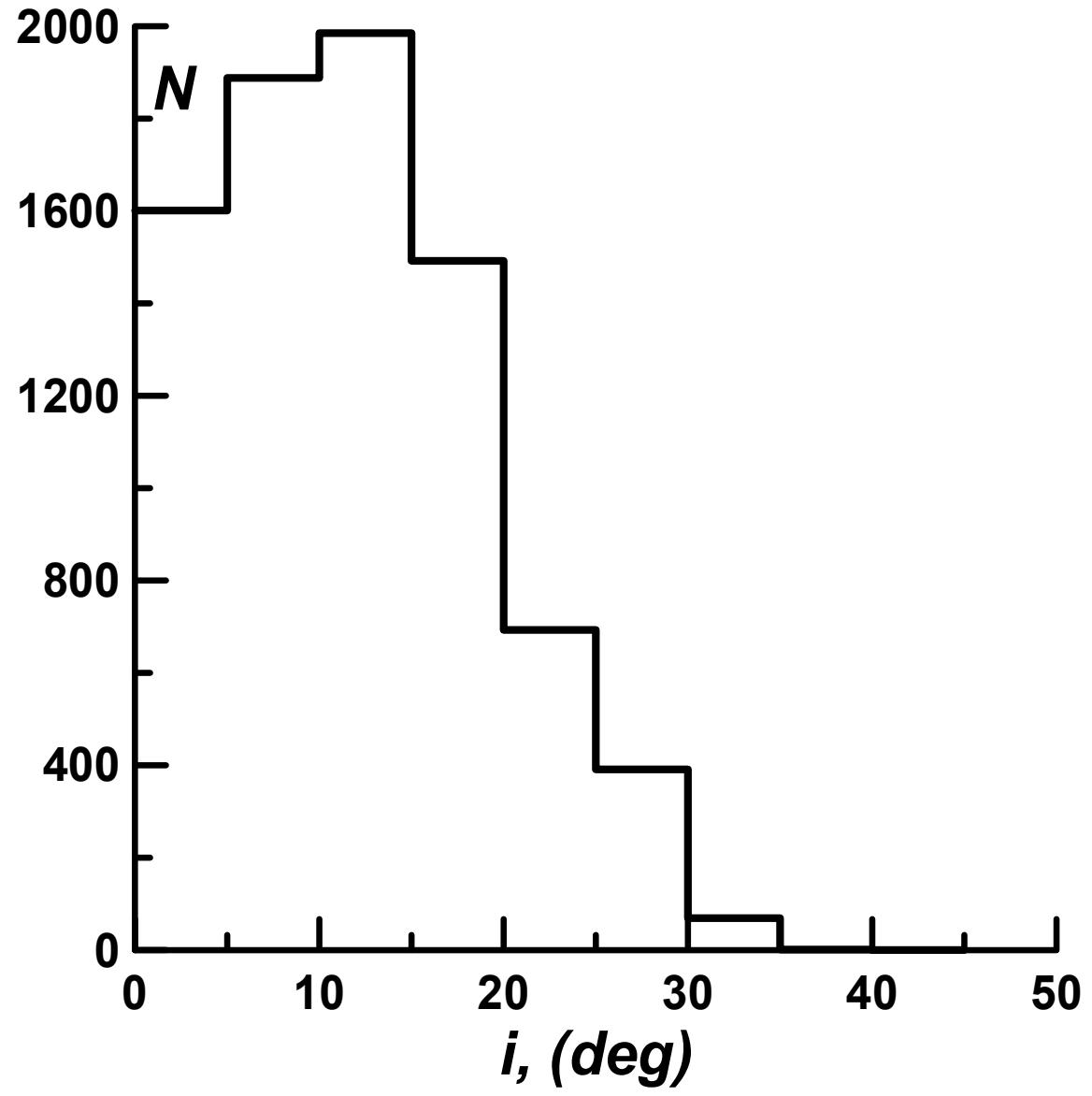

***Figure 12.*** The $N(i)$ distribution of background orbits near the 2:1 gap

The $N(i)$ distribution of the resonant orbits is presented in Fig. 11. Fig. 12 shows the $N(i)$ distribution of background orbits to the left of the resonance: $a = 3.20 - 3.25$ AU, $e = 0 - 0.30$. Absolute magnitudes of the bodies are as follows: $H < 16^m$. The total number of the orbits is 8120. A comparison of figures 11 and 12 shows that the main maximum in the distribution of resonant orbits ($i = 0^o - 5^o$) is not mainly formed by the background asteroids. This maximum could be formed in the most part by bodies of the Themis and Hygiea families.

The proper inclinations of the orbits of the bodies of the Themis family range from 0.5$^o$ to 2.8$^o$, of the Hygiea family – from 4.2$^o$ to 6.0$^o$. (Milani et al. 2014). Of course, the range of osculating elements of the orbits is somewhat wider.

The calculations show that for 112 of the 133 orbits with inclinations from 0$^o$ to 5$^o$, the inclinations themselves remain within the same limits (0$^o$ – 5$^o$) in the course of the evolution over tens of thousands years. Therefore if not all, then most of the bodies in the resonant orbits with $i < 5^o$ can quite reasonably be considered as bodies from the Themis and Hygiea families. According to Milani et al. (2014), there are no other asteroid families with proper inclinations less than 8$^o$ near the 2:1 commensurability.

The absolute magnitudes of all bodies in the resonant orbits with $i < 5^{\circ}$ are greater than $13^{m}$. Such values of $H$ correspond to sizes of the bodies of the Themis family and the Hygiea family which fell in the 2:1 MMR (Figs. 9, 10).

It is hardly to clear divide the resonant orbits of the Themis family from the orbits of the Hygiea family. But a numerical estimation one can made, using the $a' - D$ distribution of the family (Fig. 9). The figure shows that the middle of the distribution corresponds to the value of $a$ = 3.14 AU. The right edge of the distribution is truncated by the resonance at $a$ = 3.24 AU, i.e. by 0.1 AU from the centre. Assuming that the distribution is symmetric in regard to the centre, the number of bodies which fell in the resonance should be close to the number of bodies in the left wing with $a < 3.04$ AU. In this range of the $a$ values, the number of bodies with $H < 16^{m}$ in the Themis family according to Masiero et al. (2013) is 85. The similar estimation for the Hygiea family finds 26 bodies which could fall in the 2:1 resonance.

Of course, not all bodies of these families, which fell in the 2:1 commensurability zone, could go in resonant orbits. In addition, some bodies may have already left the gap. However, it is possible to draw a well-founded conclusion that the most bodies in the resonant orbits with i < $5^{\circ}$ are originated from the Themis family.

Bodies in the resonant orbits with larger inclinations can be both former background asteroids to the left of the gap and bodies of other asteroid families adjacent to the gap. One of them may be the 276 family from the list in Masiero et al. (2013).

Background asteroids are near the 3:1 and 5:2 gaps as well. In addition, there are asteroid families with wings truncated by these gaps: the 472 and 554 families near the 3:1 gap, and the 208 family near the 5:2 gap. This family in other sources (Milani et al. 2014) refers to the asteroid 158 Koronis. In Masiero et al. (2013) asteroid 158 belongs to the family as well, but in this paper its size is estimated to be slightly smaller than the size of asteroid 208. Therefore the name of the family is different. But the ranges of the proper elements of this family in Masiero et al. (2013) correspond to the ranges in other papers, in particular in Milani et al. (2014).

Thus, the background asteroids, as well as the fragments of the parent bodies of the families had to fall in each of the analyzed gaps: 3:1, 5:2 and 2:1. But for today the number of bodies in resonant orbits in the 2:1 gap is much more than in two other gaps. This difference needs to be explained.

## 4. THE MECHANISMS OF EXIT OF ASTEROIDS FROM THE GAPS

As mentioned earlier, the exit of bodies from the 3:1 and 5:2 gaps can be explained by the approaches of bodies to Mars after a sufficient increase in the eccentricities. In these resonances such an increase can occur (Wisdom 1982; Yoshikawa 1991; Moons 1996). In this study, numerical calculations of the evolution of orbits with different sets of initial elements in the 3:1 and 5:2 MMRs were also performed. In the 3:1 resonance, the eccentricities of all orbits increase from 0.05 or 0.10 to 0.32 and more at the intervals of 10,000 - 50,000 years of the evolution. When calculations are performed over 150,000-year time interval, some orbit eccentricities increase to 0.60 and even more.

In the 5:2 resonance an increase in the $e$ value from 0.05 - 0.10 to 0.40 – 0.60 occurs over a few tens of thousands years of the evolution. That is, conditions for bodies to approach Mars can appear for all orbits in both of these resonances. Since our calculations confirm the previous conclusions (Wisdom 1982; Yoshikawa 1991; Moons 1996), their detailed description can be omitted.

In the 2:1 resonance, a sufficient increase in the orbit eccentricities was not revealed (Lemaitre & Henrard 1990; Yoshikawa 1991; Moons 1996). At the same time, the gap was in reality formed, because larger bodies exist on both sides of the gap (Fig. 3). So a real mechanism of the bodies' exit from the gap should exist.

In order to a large body go out of the 2:1 gap it should also approach a planet. And it is possible only for the bodies in orbits with significant eccentricities. According to our calculations, the eccentricity of the Mars orbit changes from 0.066 to 0.111 with a period about 50,000 yr, and the eccentricity of the Jupiter orbit changes from 0.026 to 0.061 with a period about 30,000 yr. So the maximum aphelion distance of the Mars orbit is 1.69 AU, and the minimum perihelion distance of Jupiter orbit is 4.88 AU. Thus, in order to a body approaches Mars, it should be in the orbit with $e > 0.48$. The asteroid approaches Jupiter are possible at the same $e$ values.

One can see from Fig. 7, there are orbits with rather significant eccentricities among the resonant orbits. The eccentricities of 8 orbits exceed 0.50. That is, bodies in these orbits can approach the planets and go out of the resonance. In our previous study (Kazantsev & Kazantseva 2021) six asteroids with $H < 18^m$ are found to have go out of the 2:1 resonance to the Centaur population in the next few thousand years or several tens of thousands years.

Numerical calculations were performed to evaluate the evolution of the orbits in the resonance with different initial elements. Perturbations from all eight major planets (Mercury - Neptune) were taken into account, as well as the relativistic effects of orbital perihelion displacement.

In the course of the orbit evolutions over time interval of the hundreds of thousands years, no increases in the eccentricities of the orbits from 0.05, 0.10 and 0.20 to values sufficient for bodies to approach the planets were detected. These calculations confirm the results of the previous studies (Lemaitre & Henrard 1990; Yoshikawa 1991; Moons 1996). Therefore, there is no need to provide a detailed analysis of them.

These calculations explain the significant number of smaller bodies in the 2:1 resonance and their almost complete absence in the 3:1 and 5:2 resonances. But it is still unclear the absence of larger bodies ($H < 12^m$) in all three resonances (Figs 1-3).

It is clear that the lifetime of smaller bodies in the gaps can only be shorter. Therefore, the bodies which are in the gaps today (i.e. in the resonant orbits), have got there much later than the gaps themselves were formed. Once, rather large bodies had come out of the gaps. The exit of bodies from the gaps in the 3: 1 and 5: 2 resonances can be explained by approaches to Mars. And what is the mechanism of exit of bodies from the gap in the 2: 1 resonance?

Our calculations of the evolution of the orbits in the 2: 1 resonance are performed taking into account perturbations from all major planets, as well as the relativistic effects of orbital perihelion displacement. And these calculations, as well as the calculations of other authors (Lemaitre & Henrard 1990; Yoshikawa 1991; Moons 1996), did not show a sufficient increase in the eccentricities. Probably in reality there are some factors that were not taken into account in our calculations and in other studies, but which cause a noticeable increase in the eccentricities of the resonant orbits. In our opinion, such factors may be non-gravitational effects (NGEs). It is difficult to offer another explanation. Most probably, the Yarkovsky effect may be such NGE. Firstly, it was already found an increase in the eccentricities of the resonant orbits under the action of the Yarkovsky effect (Brož & Vokrouhlický 2008; Xu, Zhou & Ip 2020).

Secondly, in Chrenko et al. (2015) a possibility of exit of bodies from the 2:1 resonance, when the Yarkovsky effect is taken into account, was found. However, in that paper, the orbit evolution of relatively small bodies (up to 18 km) was calculated. According to Chrenko et al. (2015), the exit time of bodies up to 5 km in size should be from several hundreds of millions years to one billion years. Since the additional momentum of the body due to the Yarkovsky effect $dp$ is approximately inversely proportional to its size, the exit time from the gap of bodies of 50 km in size should be about 10 billion years. So it turns out that even taking into account the Yarkovsky effect cannot explain the exit of large bodies from the 2:1 gap.

In Chrenko et al. (2015) calculations over intervals of several billions of years were performed. It is difficult to determine the accuracy of the calculations over such long intervals. In the present study, another approach was used to estimation the influence of the Yarkovsky effect on the exit of bodies from the 2:1 gap. The calculations were performed up to much shorter intervals, where the accuracy of the calculations can still be considered as satisfactory.

## 5. A POSSIBLE INFLUENCE OF NGEs ON THE EVOLUTION OF THE RESONANT ORBITS

### 5.1 The influence of a model NGE

The essence of the Yarkovsky effect is that a body obtains an additional impulse *dp* due to the re-radiation (in the infrared range) of the absorbed sunlight. There is a certain delay *dt* between the moment of the sunlight absorption and its re-radiation. Therefore, an additional velocity *dv* to the heliocentric orbital velocity of the body *vh* arises.

The increase in heliocentric velocity depends not only on the mass of the body, but also on the orbit elements of the asteroid, on the albedo of its surface, on the spin vector orientation and spin period.

For a body in a circular orbit, the maximum of the additional velocity will be in case of the spin axis of the body is perpendicular to the plane of its orbit, and the value of *dt* is a quarter of its spin period. With prograde rotation the additional velocity will be positive, and with retrograde rotation - negative.

The action of the Yarkovsky effect is too small to be noticeable in the orbit evolution of an asteroid of several kilometers in size at time intervals of thousands or tens of thousands years. Therefore, numerical calculations of the evolution of asteroid orbits taking into account the action of a model NGE were performed. The essence of the model NGE action corresponds to the action of the real Yarkovsky effect. In the program one can change the asteroid albedo, the spin vector orientation and the spin period. Only the value of the additional impulse was chosen much larger than in case of the real Yarkovsky effect.

Numerical calculations of the evolution of the orbits of bodies in the 2:1 resonance with artificial additional impulses in a wide range were performed. In case of the spin axis of the body was chosen perpendicular to the plane of its orbit, the range of values of the additional velocity *dv* ranged from 0.01 mm/s to 5 mm/s per day.

The *dv* value was adding to the *vh* value in each integration step. Axial rotation of the body was chosen both prograde ($dv > 0$) and retrograde ($dv < 0$). It turned out that at all positive values of *dv* a continual increase in the eccentricities of the resonant orbits take place, in contrast to the case of $dv = 0$ (Fig. 13).

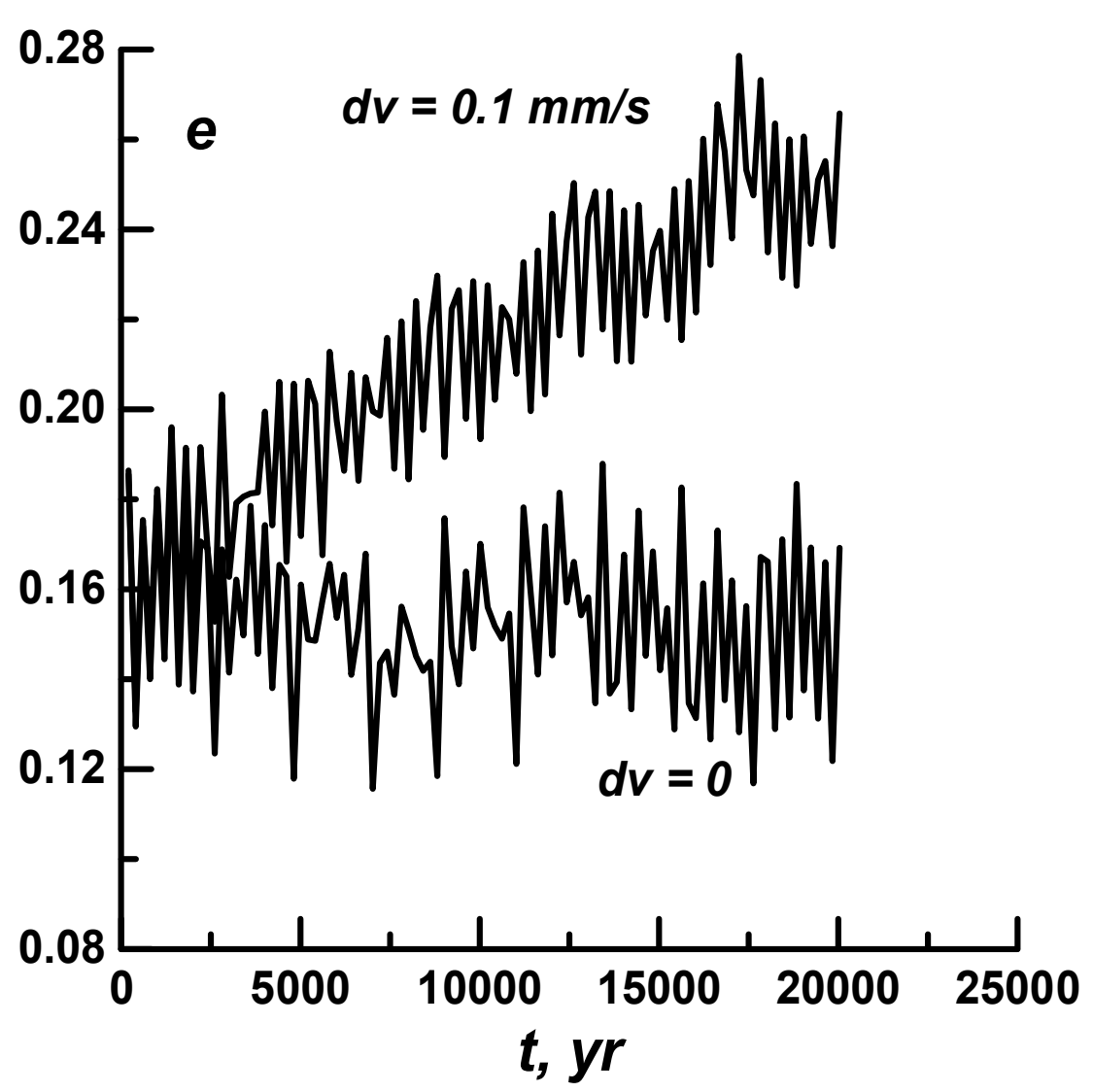

***Figure 13.*** The *e*(*t*) dependences of an orbit with action of the NGE and without it

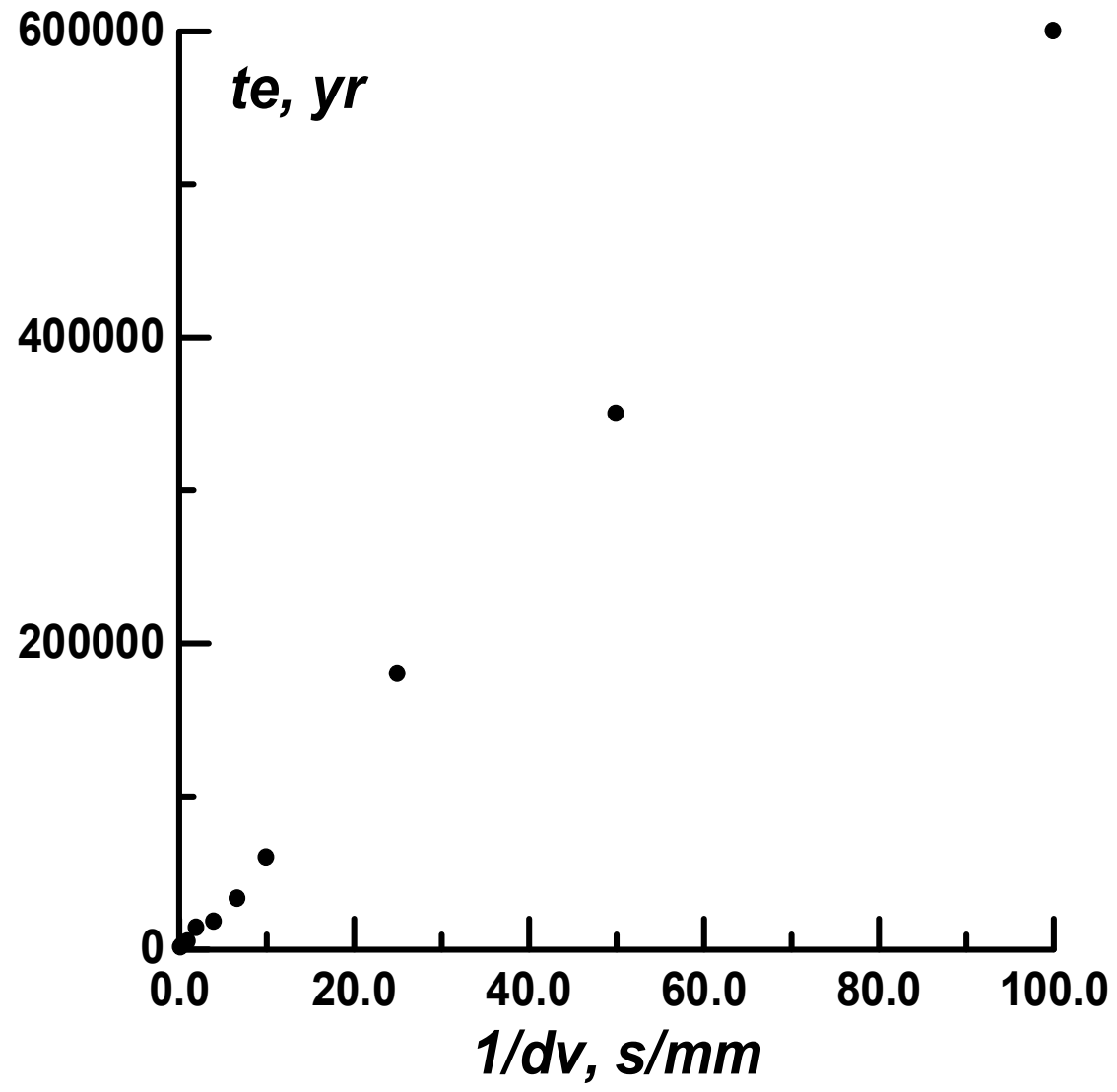

***Figure 14.*** The average time of increase in the eccentricities on the 1/*dv* value

The average time of increase in eccentricities *te* almost linearly depends on the value of 1/*dv* (Fig. 14).

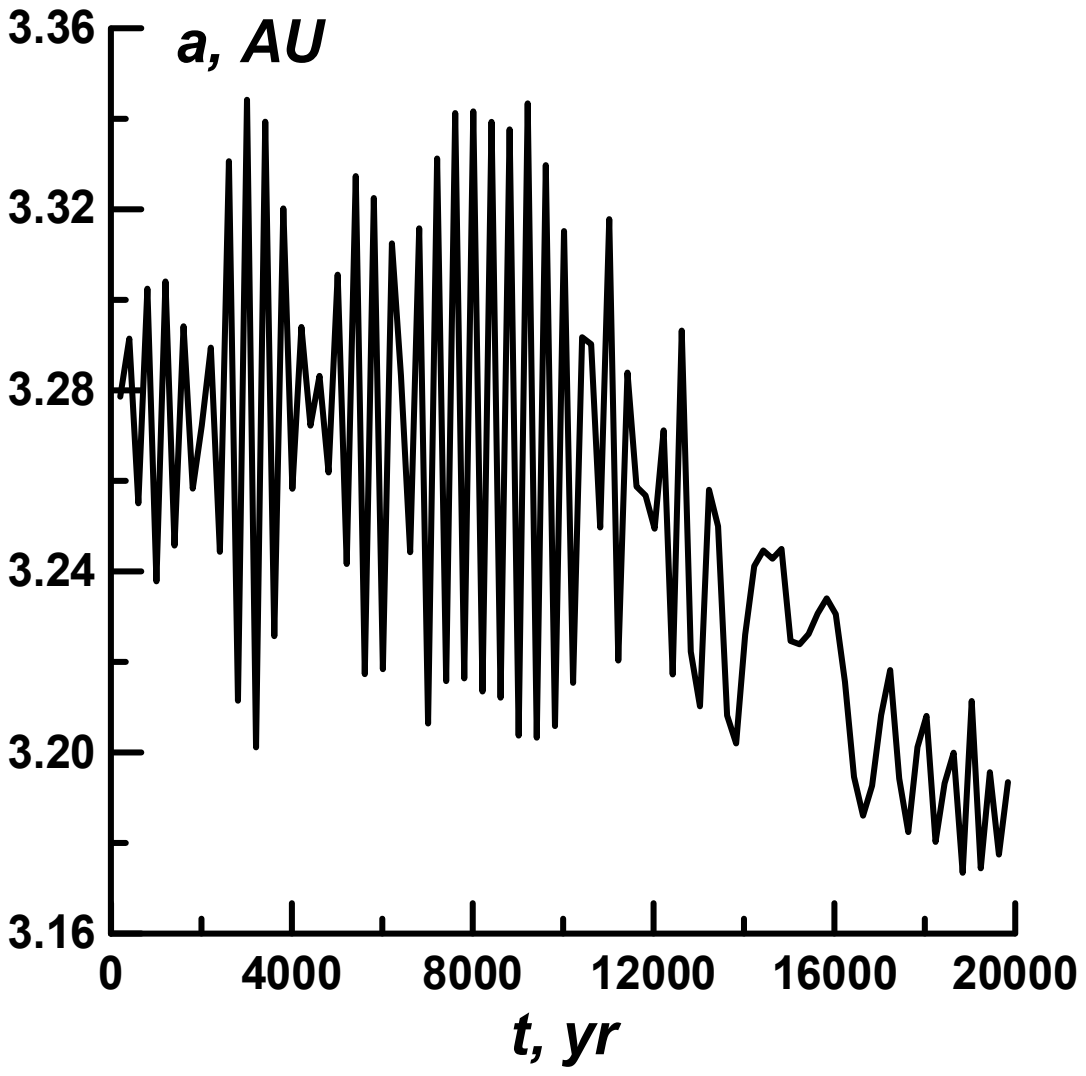

***Figure 15.*** The *a(t) dependence* of an orbit with $dv = -0.1$ mm/s per day

At the minimum $dv$ values (0.01 mm/s per day) the time of increase in eccentricities from 0.15 to 0.50 is about 600,000 yr, at the maximum (5 mm/s per day) – about 1,300 yr. It is not advisable to perform the calculations with a very little $dv$ value, because the calculation intervals will be much more than one million years. The calculation accuracy is very low at such intervals.

At negative, even small values of $dv$, the orbit exits from the resonance in the zone of the semimajor axes $a < a_c$ (Fig. 15). Therefore, all bodies should exit from the resonance under the action of NGE, regardless of the value and the direction of the additional velocity.

### 5.2 The influence of the Yarkovsky effect on the exit of asteroids from the gap

An influence of the Yarkovsky effect on the exit of a model body of 5 km in size from the 2:1 resonance was estimated. This size is near to the average size of the selected asteroids in orbits with $e > 0.5$ in the resonance. Absolute magnitudes of these asteroids range from $14^m$ to $16^m$. The albedo $p$ and the bulk density ρ of the model asteroid were adopted as follows: $p = 0.05$, $\rho = 1$ g/sm$^3$.

The amount of the solar radiation that falls on the surface of the asteroid per a unit of time at a distance of 3.30 AU from the Sun can be estimated by the solar constant. If all the radiation, which is absorbed by the asteroid, will be emitted in the direction opposite to the direction of the **vh** vector, the additional velocity received by the body will be approximately equal $1\times10^{-5}$ mm/s per day. If the above obtained $te(dv)$ dependence is maintaining up to such values of $dv$, the time of increase in the eccentricity of the orbit from 0.15 to 0.5 will be about 600 million years.

This estimate is obtained on the assumption that all quanta are re-emitted in the direction opposite to the direction of the **vh** vector. In reality, these quanta should re-radiate over the entire hemisphere. Assuming uniform distribution of the radiation directions, the time of the increase in the eccentricity should be about 1 billion years.

However, this is also the minimum time. After all, it is obtained in the case of the most favourable orientation of the re-radiating hemisphere relatively to the direction of the **vh** vector. It is clear that such an orientation is quite rare. Therefore in most cases the $te$ value for a body of 5 km in size should be noticeable greater than 1 billion years.

The obtained here estimation of the exit time from the 2:1 gap of a body of 5 km in size is quite close to the corresponding estimation obtained in Chrenko et al. (2015), where a different approach to calculations was used. But the average densities of the bodies were taken the same. Therefore, it can be assumed that the exit time from the 2:1 gap of bodies of about 5 km in size, taking into account the Yarkovsky effect, should really be about 1 billion years or even more.

It is difficult to estimate the lifetime of each orbit in the 2:1 resonance. One can only assume that orbits with large eccentricities are in the resonance longer than others.

## 6. THE PROBLEM OF THE ORIGIN OF THE GAP IN THE 2:1 RESONANCE

Now it is worth to pay attention to another problem of the gap in the 2:1 resonance: how to explain the formation of the gap in an earlier era? One can see from Fig. 3 that the gap really exists for bodies with $H < 12^m$. The number of bodies on the left and on the right of the gap is markedly different. The bodies on the left could move there in some way later from the more inner zones of the asteroid belt, but the bodies on the right had to exist there from the very beginning, i.e., not latter than the gap was formed. Therefore, the concentration of bodies in the gap at the beginning had to be not less than the concentration of bodies on the right. That is, about several dozen of bodies with $H < 12^m$ (tens of kilometres in sizes) had to be in the gap. And almost all of them left the gap before the debris of neighbouring families had fallen there. This is the only a way to explain the absence of larger bodies in the gap and the presence of smaller ones.

The dependence of the $dv$ value on the size $D$ of a body due to the Yarkovsky effect is approximately as follows: $dv = 1/D$. The exit time of bodies of 5 km in size from the gap is not less than 1 billion years Therefore, the exit time of bodies of 50 km in size from the gap due to action of the Yarkovsky effect should be tens of billions years.

How can such an obvious paradox be explained? Probably only in the manner, that in the early Solar system, the physical conditions for bodies in the asteroid belt were significantly different from the today ones. As it can be seen from Figs 1-3, the location of the gaps for larger asteroids corresponds to the to-day value of the semimajor axis of the Jupiter orbit. Therefore, the gravitational influence from the major planets on the asteroids in the 3:1, 5:2 and 2:1 MMRs in the early epoch had to be similar to the today influence.

Thus, it remains to assume that namely the NGEs influenced on the bodies in the asteroid belt in the early Solar system a much stronger in compare with the today NGEs. One of the factors that contributed to a stronger manifestation of the NGEs in the early Solar system could be a significantly higher intensity of the solar radiation, as compared to the current one.

The age of the Themis family is estimated of from 2.4 to 3.8 billion years (Spoto, Milani & Kneˇzevic 2015). A very high intensity of the solar radiation had to end before the Themis family arose, that is, earlier than 2.5 billion years ago. Otherwise, members of this family had not only to fall in the 2:1 gap, but even had to go out of it completely.

In some papers (Feigelson, Garmire & Pravdo 2002; Mishra & Marhas 2019) it was pointed on a possibility of more powerful the solar radiation in the distant past. In particular, it is noted in Feigelson, Garmire & Pravdo (2002) that X-ray bursts on the young Sun could occur hundreds times more frequent and tens times more powerful compared to the contemporary level. And the flow of energetic protons from the early Sun was $10^5$ times greater.

Thus, the assumption made here about a much higher intensity of the solar radiation in the distant past has some independent confirmation. If this assumption will receive the final confirmation with clear quantitative characteristics, it will, among other things, close the question of the origin of the Kirkwood gaps in the asteroid belt.

## 7. **CONCLUSIONS**

For all orbits in the 3:1 and 5:2 MMRs an increase in the orbit eccentricities, which are sufficient for the bodies to approach Mars, was obtained. In the 2:1 MMR, a sufficient increase in the orbit eccentricities was not detected.

The debris of the parent bodies of the asteroid families adjacent to the gap, in particular, the Themis family, consist a noticeable part of the existing bodies in the 2:1 resonance.

A significant increase in the eccentricities of the orbits in the 2:1 resonance can be explained by the action of the Yarkovsky effect. The action of the Yarkovsky effect can explain the exit of an asteroid with a size of 5 km from the 2:1 MMR over a period about 1 billion years or more.

More than 2 billion years ago, there were dozens of bodies over 50 km in size in the 2:1 gap. To form the gap in the 2:1 resonance at the very beginning, the physical conditions in the asteroid belt had to be significantly different from the today ones. In particular, the intensity of the solar radiation in the early Solar system could be much higher as compared to the today one.